\documentclass[pra,aps,preprint]{revtex4}
\usepackage{amsmath,graphicx,epsfig}

\newlength{\defbaselineskip}
\newcommand{\setlinespacing}[1]%
           {\setlength{\baselineskip}{#1 \defbaselineskip}}

\begin{document}

\title{Hybrid ion, atom and light trap\\}
\author{S. Jyothi, Tridib Ray, N. Bhargava Ram, S. A. Rangwala}
\affiliation{Light and Matter Physics, Raman Research Institute, Bangalore 560080, India}



\begin{abstract}

We present an unique experimental arrangement which permits the simultaneous trapping and cooling of ions and neutral atoms, within a Fabry-Perot (FP) cavity. The versatility of this hybrid trap experiment enables a variety of studies with trapped mixtures. The motivations behind the production of such a hybrid trap system are explained, followed by details of how the experiment is put together. Several experiments that have been performed with this system are presented and some opportunities with this system are discussed. However the primary emphasis is focussed on the aspects that pertain to the trapped ions, in this hybrid system. 

\end{abstract}
\date{\today}
\maketitle
\section{Introduction}
\label{intro}
Cooled and trapped atoms and ions have individually been exceptionally important for experiments, ranging from precision measurements~\cite{Ye2008,Blaum2010} to studies of few and many particle physics~\cite{Leggett2001}.
This is due to our ability to create these systems in specific motional and internal quantum states, with long trapping times and cold temperatures, allowing precise interrogation of resonances and system response to pertubations. Extending such techniques to trapped mixtures allows inter-species interactions to be measured with most of the above advantages and opens up new possibilities for the study of few and many particle systems. However, usually precision and mixtures are incompatible (because the complexity and demands increase) and special care needs to be taken to retain the advantages of precise manipulation and interrogation while allowing individual components (ions and atoms) to cohabit simultaneously. A good hybrid trap permits the study of such mixtures, with minimum disruption due to the multiple technologies employed to create, manipulate the mixture and detect the resulting phenomena.

Ion-neutral interactions at collision energies below 1 eV is largely unexplored. This is because such experiments have either been done with beams or trapped ions confined within a uniform density distribution of neutral gas~\cite{Church1993}. Developments in the cooling of atoms~\cite{Chu1985,Raab1987,Anderson1995,Davis1995} and ions~\cite{Neuhauser1978,Wineland1978} have enabled new experimental systems to be realized, which in effect can allow the binary collision energy for ions and atoms to from eV to $\mu$eV to be investigated. In addition the strong attractive ion-atom interaction ($-\alpha/r^4$, where $\alpha$ is the atomic polarizability and $r$ is the ion-atom seperation) can lead to few and many particle interactions in the dilute gas mixture~\cite{Cote2000}. Since the trapping mechanisms for ions and atoms are quite different, the construction of hybrid traps capable of holding cold ensembles of both is not obvious. However, recently several groups have successfully demonstrated hybrid traps for this purpose, by combining magneto-optically trapped atoms~\cite{Smith2005,Grier2009,Ravi2012a,Hall2012}, or degenerate atomic gases~\cite{Zipkes2010,Schmid2010} with ion(s) in versions of the linear Paul trap.

In this article we discuss a hybrid trap for cold atoms overlapped with ions within an trap analogus to a spherical Paul trap~\cite{Paul1953}. The experimental configuration allows the construction of a moderate finesse Fabry-Perot (FP) cavity around the trapped ion-atom mixture.  This addition can be viewed as the next level of hybridization, where the atoms in the cavity mode can exhibit collective strong coupling with cavity photons on atomic resonances, therefore opening up fresh possibilities to manipulate and detect ion-atom mixtures. Due to the constraints inherent in the construction of the experiment, the ion trap constructed here has several non-standard features. Below we discuss these features in some detail, highlighting the advantages of this construction, presenting some experimental results and future prospects. 

\section{Hybrid trap assembly}
\subsection{Vacuum system and electrode configuration}
\begin{figure}[t]
\centering
\includegraphics[width=15 cm]{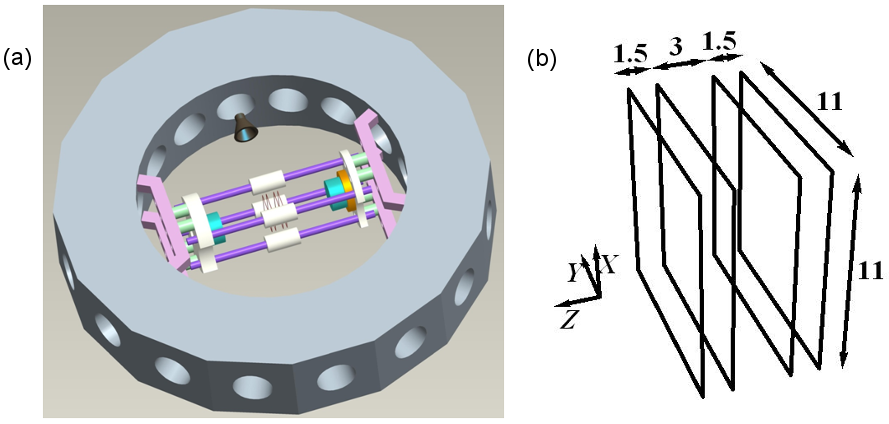}\\
\caption{(a) Shows the schematic of the experimental set up. The four quadrupolar rod structure is shown and the cavity is constructed along the quadrupolar axis. The thin-wire ion trap is wound around the ceramic sleeves in the central region of the rod structure. The position of the CEM is shown. (b) shows the four wire configuration with dimensions in mm.}
\label{fig:trapconfig}
\end{figure}

The schematic diagram for ultra high vacuum chamber ($\approx10^{-10}$ mbar) and the principle components of the experiment is illustrated in Fig.~\ref{fig:trapconfig}(a). The experiment is contained within a compact stainless steel chamber (Kimball Physics{\texttrademark}) with 16 CF16 and 2 CF100 ports, which allows sufficient optical and electrical access. The electrode configuration in Fig.~\ref{fig:trapconfig}(a) allows both a linear and a spherical Paul trap to be operated (though not simultaneously). The central electrode assembly has 4 stainless steel rods of 3 mm diameter held in a quadrupolar configuration, with center to center distance of 17 mm. The rod assembly is arranged symmetrically about the chamber diameter that runs through the center of two of the 16, CF16 ports of the chamber. Periodically grooved ceramic sleaves are attached to the rods at the center of the chamber, and around these sleeves four 80 $\mu$m tungsten wires are tightly wound in identical square configuration, separated by $1.5$ mm, $3$ mm and $1.5$ mm respectively (Fig.~\ref{fig:trapconfig}(b)), so as to define the thin-wire electrodes which constitute the modified spherical Paul trap (modified spherical is used to reflect the seriously distorted electrode geometry from the hyperbolic electrode shape of an ideal Paul trap, while keeping the operating principle the same). Application of rf voltage on the central two wires allows the trapping of ions, while a small constant voltage on the peripheral electrodes permits control over the trap distortion and manipulation of the trap center. Fig.~\ref{fig:IonTrap} (a) and (b) shows the electric field in the XY and XZ planes when the rf phase is at the positive maxima. Alternatively, out of phase rf voltages on diagonally paired rod electrodes, while using the wires electrodes as endcaps, makes a linear Paul trap. Here we shall only discuss the operation of the spherical trap.
\renewcommand{\figurename}{Figure}
\subsection{Atom trap(s)} Contained within the experiment are rubidium (Rb), potassium (K) and cesium (Cs) dispensers, any combination of atoms from which can be cooled and trapped in a magneto-optical trap (MOT). Two coils in anti-Helmholtz configuration, placed outside the CF100 viewports provide the gradient magnetic field essential for the MOT for the above atomic species. Appropriate cooling and repumping lasers, comprising three pairs of orthogonal beams, are utilized to laser cool the atoms and there is sufficient flexibility to form MOT's of different spatial sizes and densities. The design of the instrument ensures that the shadow of the wires avoids the geometric centre of the chamber and therefore does not affect the MOT formation. A typical MOT profile, measured in fluorescence is shown in Fig.~\ref{fig:AtomCavity}(a). Details of the MOT production and diagnostic have been discussed in earlier work~\cite{Ray2013a,Ray2013b}.

\subsection{Modified spherical Paul trap} 
\begin{figure}[t]
\centering
\includegraphics[width=15 cm]{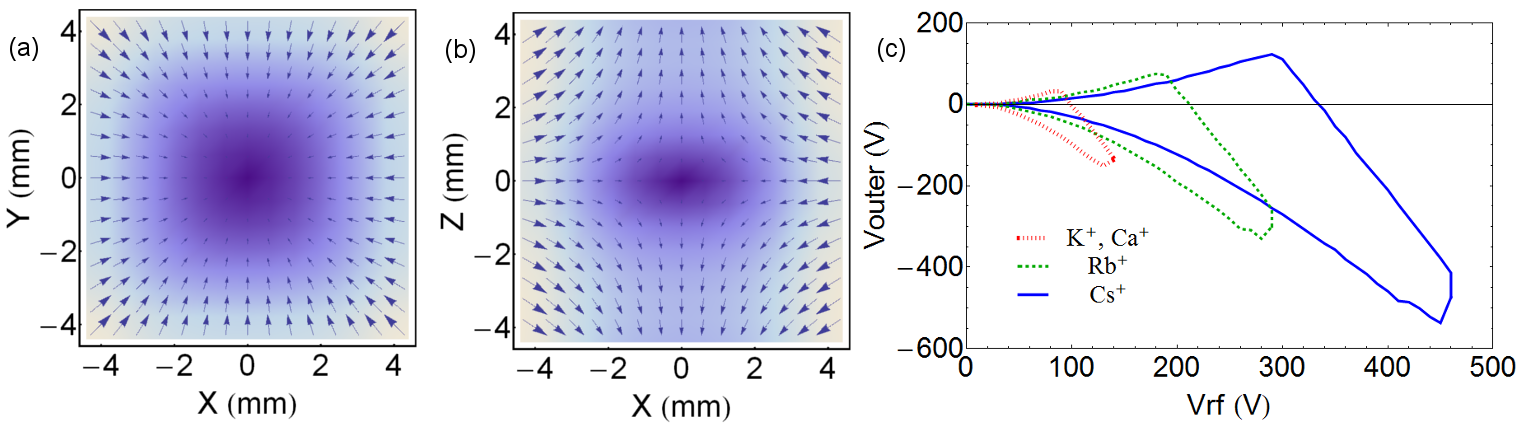}
\caption{(a) and (b) shows the electric fields in the XY and XZ planes passing through the trap center when the inner pair of wires are at a positive voltage. In both figures, light to dark backgrounds indicates the  magnitudes of the quantities being plotted, from large to small. The secular trap is formed when an rf is applied on the inner wire pair. (c) shows the region of stability as a function of applide voltages for K$^{+}$, Ca$^{+}$, Rb$^{+}$ and Cs$^{+}$ ions at rf frequency 500 kHz, where V$_{outer}$ is the constant voltage on the outer two wires and V$_{rf}$ is the amplitude of the rf voltage on the inner two wires.}
\label{fig:IonTrap}
\end{figure}

The thin-wire ion trap (T-WIT) is loaded with atomic ions by two-photon ionization from the MOT. For the case of the $^{85}$Rb$^+$ ion, we find~\cite{Ray2013b} that optimal stable ion trapping occurs at an rf voltage, $V_{rf}=80$ V, and a constant voltage, $V_{outer}=-5$ V on the outer pair of wires, when the rf value is $\nu_{rf}=500$ kHz. The secular trap depth is determined to be $~300$ meV for $^{85}$Rb$^+$ trapped in the field configuration mentioned above. In addition to the alkali atom sources, a calcium (Ca) dispenser is also present, which can once again be two-photon ionized from  hot vapour and trapped. The stable operating voltage combinations ($V_{rf}$ on inner wire pair and $V_{outer}$ on the outer wire pair) is found by solving the single particle equation of motion in Mathematica{\texttrademark} using potentials derived in SIMION{\texttrademark}. Fig.~\ref{fig:IonTrap}(c) shows the stability region for K$^{+}$, Ca$^{+}$, Rb$^{+}$ and Cs$^{+}$ ions at rf frequency 500 kHz. The Ca$^+$ ions can be directly laser cooled and therefore can also be detected by fluorescence. 
\subsection{Ion detection} As the ionized alkali atoms do not fluoresce at optical frequencies due to their closed shell electronic structure, alternative methods for detection are required.  In the experiment, the channel electron multiplier (CEM), operates as a charge counting device. The natural location for the CEM is on the symmetry axis of the system~\cite{Ravi2012a}. Since that is also the cavity axis, the CEM is mounted transverse to the axis of the experiment, as illustrated in Fig.~\ref{fig:trapconfig}. Here only the schematic assembly is illustrated, without the details of the CEM housing and the isolation grid in front of the CEM, to shield the trapped ions from the very large voltages applied to the CEM. The off-axis transverse extraction from the ion trap is a complex process, and the reliability needs to be established, in terms of the ion numbers detected, so that the number of ions present in the atom-cavity system at any given time can be measured. Due to the various voltages present in the system, pulsed extraction of the ions from the trap is implemented. As seen in Fig.~\ref{fig:trapconfig}, the proximity of the CEM coupled with large extraction voltages result in a few microseconds of flight time and the arrival time distribution can be very small, leading to detection pile-up (detector saturation). Strategies have therefore been evolved measure a large number of trapped ions with confidence in our experiment~\cite{Jyothi2013}. 

\subsection{Fabry-Perot cavity} The centers of the atom trap and the ion trap coincides with the center of the chamber, which is on the axis of the Fabry- Perot (FP) cavity. This is ensured by referencing all the traps with the four SS316 rod structure and therefore constructing about a common axis. Small adjustments in the position of the trapped atom and ion distributions are possible to optimize the physical overlap of these with each other and the cavity mode. The FP cavity consists of two 
mirrors of radius of curvature $~50$ mm, mounted at a separation of $~45.7$ mm in a symmetric, near confocal geometry. One of the mirrors is mounted on a piezoelectric stack, in order to tune the cavity length by application a voltage. Shielding electrodes are put in place (not shown in the schematic) so that the effect of the tuning voltage for the piezo can be corrected at the location of the ion trap. The spatial profile of the cavity mode is imaged on a CCD camera and the transmitted light intensity is measured on a photo multiplier tube (PMT). The cavity linewidth is measured to be $1.4\pm (0.2)$ MHz at $780$ nm, corresponding to a finesse of $\approx 2400$. The waist at the centre of the TEM$_{00}$ mode is 78 $\mu$m.

\section{Experiments with the hybrid trap}

\subsection{Ion-atom mixture experiments} The above instrument has been used to demonstrate simultaneous trapping of Rb atoms, and Rb$^+$ ions~\cite{Ray2013b}.  The ions are created by the resonant two-photon ionization of MOT atoms~\cite{Ravi2012a} (780nm + 473nm), in close proximity to the center of the overlapped ion trap, and with the velocity of the parent atom. This ensures that the loading efficiency of the ions is very high, because matching the ion to the instantaneous phase of the rf field is automatically satisfied. However the trapping field pumps kinetic energy into the ions within a few rf cycles, so the initially small ion velocities rapidly increase to reflect the trap depth. In this circumstance, the ions need to be cooled so that they can be trapped for long times and their populations in the ion trap is stabilized. Since for alkali ions the lowest energy fluorescing transitions lie in the vacuum ultraviolet, straight-forward laser cooling is not possible. Using such wavelengths for laser cooling would ionize the MOT atoms~\cite{CRC2004} and therefore not be conducive for producing a stable ion-atom mixture. \\
\begin{figure}[t]
\centering
\includegraphics[width=15 cm]{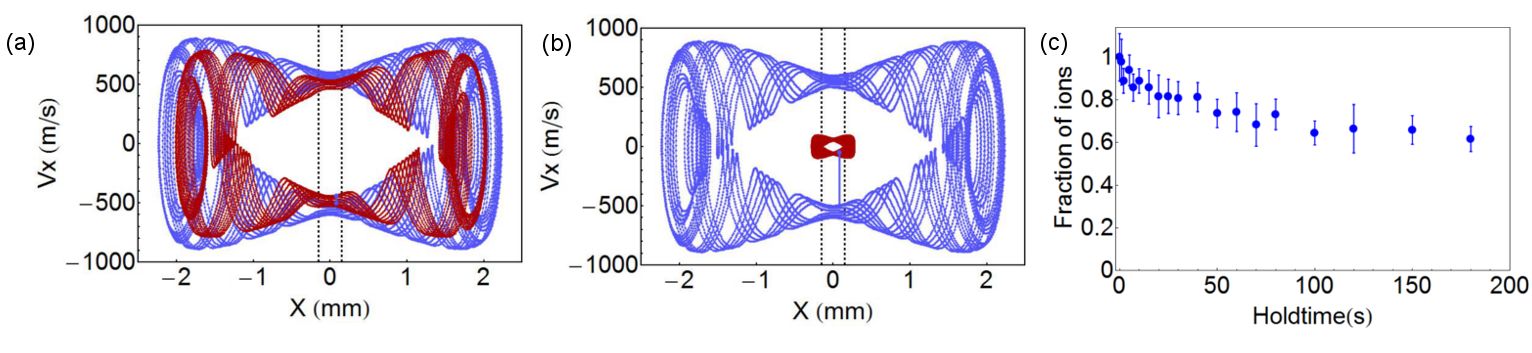}
\caption{(a) and (b) Shows the phase space diagram of  $^{85}$Rb$^{+}$ ion before collision(blue) and post collision(red) with its parent atom. (a) represents an elastic and (b) resonant charge exchange collision respectively. The dashed line shows the typical FWHM of the MOT. (c) shows the stabilization of $^{85}$Rb$^{+}$ ion when held with the cold $^{85}$Rb atoms as discussed in section 3.1.}
\label{fig:IonAtom}
\end{figure}
The above experimental situation differs significantly from when the ions and atoms are independently laser cooled and mixed, because in that case, the steady state engineered is only very weakly (if at all) dependent on the ion-atom interaction. Here however, no independent cooling channel exists for the trapped ions and therefore, if the trapped ions are to survive, the interaction (binary ion-atom collision) must cool them. This has been demonstrated in our experiments with Rb mixtures~\cite{Ray2013b,Ravi2012b}, and independently with Na mixtures~\cite{Sivarajah2012}. These experiments therefore show that sympathetic cooling by collisions between ions and atoms of equal mass is possible and efficient. At the collision energies in the experiment, elastic and resonant charge exchange collisions are the dominant binary collision channels. Since the atoms are laser cooled, they can be assumed to be at rest, relative to the velocities of the ion. Therefore, in any ion-atom collision, the atom gains kinetic energy. Whether the ion cools in the collision depends on whether the micromotion (field synchronous motion), or the macromotion (secular orbital motion) energy is transferred to the atom. Since the micromotion is ion position and rf field dependent, any collisional reduction in micromotion velocity is regained instantly post collision, and therefore does not contribute to ion cooling. A collisional reduction of macromotion however results in a tighter ion orbit, and therefore a cooling of ion motion. Since the macromotion velocity is maximum at the center of the ion trap, which is the location of the MOT atoms (the MOT volume is much smaller than the ion trap volume), the conditions are ideal for collisional cooling of the ion. While both elastic and resonant charge exchange collisions participate in the cooling of the ion and it is the localized atom distribution which results in collisions exclusively at the trap center that enables the cooling~\cite{Ravi2012b}. Further, in the resonant charge exchange channel, the prospect of bringing a fast ion to rest, in a single glancing collision exists, when the ion and the atom swap charge states, without significant change in the dynamical quantities. Fig.~\ref{fig:IonAtom}(a) and (b) illustrates elastic and resonant charge exchange collisions respectively. Here the phase space of an ion before collision and post collision are shown. The angle of deflection for the ion in the elastic collision case is 10$^\circ$. Both elastic and the charge exchange collision occur at the same spatial point, but the glancing resonant charge exchange is much better at cooling the hot ion. The closer to the center the collision occurs, the cooler the ion. The parent atom - daughter ion system therefore provides a new method for cooling fast ions in a single, glancing collision. In Fig.~\ref{fig:IonAtom}(c), we show the extremely long lifetimes of ions, when approximately 100 ions are loaded into the ion trap from the MOT, and roughly 70\% of them survive for very long hold times, when held in contact with the parent MOT in the combined trap discussed above. New methods have been developed to measure the collision rate coefficients between the atoms and the trapped ions by monitoring the changes in the atomic fluorescence~\cite{Lee2013}.

\subsection{Requirement for a cavity} 
\begin{figure}[t]
\centering
\includegraphics[width=15 cm]{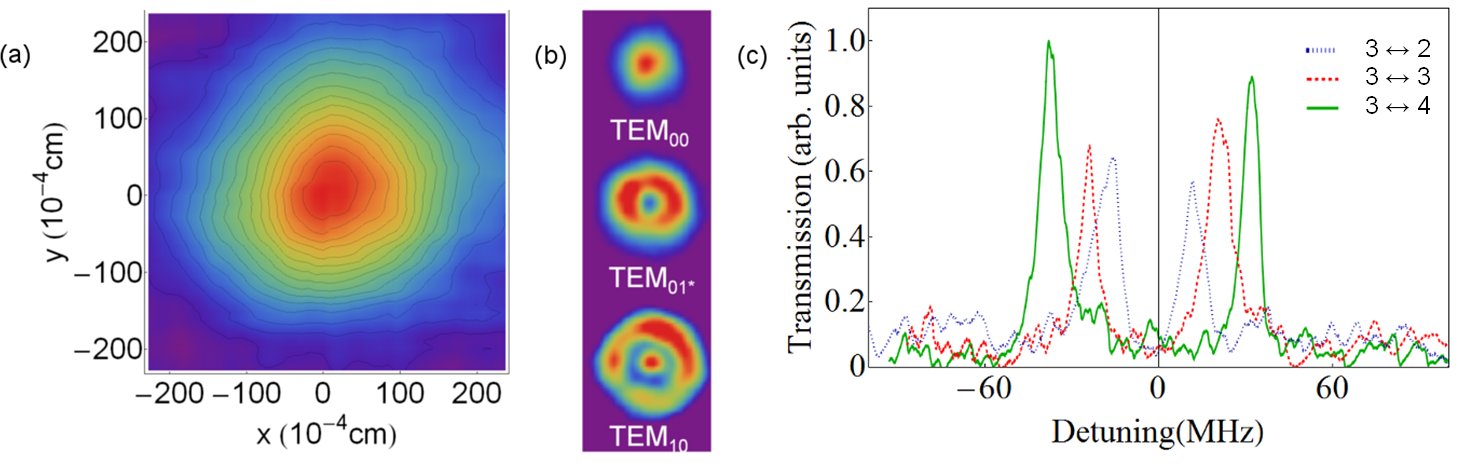}
\caption{(a) shows the spatial profile of a typical MOT. (b) shows the TEM$_{00}$, TEM$_{01^{*}}$ and TEM$_{10}$ modes coupled to the cavity from the MOT fluorescence. In the false colour images (a) and (b), red indicates high light intensity and purple no light intensity. (c) shows the collective normal mode splitting on three different atomic transitions of $^{85}$Rb as discussed in Section 3.3.}
\label{fig:AtomCavity}
\end{figure}

The purpose behind building the cavity around the ion-atom mixtures is (a) to use the cavity to manipulate the cold trapped species and (b) for probing of the results of ion-atom interactions, in-situ. Manipulation of the trapped system can be done, for example, by imposing a standing wave dipole trapping field on the mixture. Typically, the laser cooled species can be detected in fluorescence. However, when multiple species are present, then the ability to detect a specific species or state becomes very challenging. The FP cavity such as the one discussed above, couples only to specific frequencies and therefore is a very high quality spectral filter for the detection. Detection can either be by fluorescence coupling from the trapped quantum system (Fig.~\ref{fig:AtomCavity}(b)) or by developing the atom-cavity collective strong coupling (Fig.~\ref{fig:AtomCavity}(c)) as a generic tool for the detection of interactions as is discussed below. 
\subsection{Atom-cavity collective strong coupling experiment}
In the dressed state picture, a single atom in its ground state with a photon in the single electromagnetic(EM) mode, $\left|g,1\right\rangle$, is degenerate with the atom in its exited state with no photon in the mode, $\left|e,0\right\rangle$~\cite{Berman1994,Haroche2006}. When the atom couples strongly with the EM mode, as a result of the interaction, the degeneracy is lifted by a factor of 2g$_{0}$, where g$_{0}$ is the single atom-cavity coupling~\cite{Berman1994,Haroche2006,Jaynes1963}. The strong coupling is manifest when $g_0 >$($\kappa,\gamma$), where kappa is the photon decay rate from the cavity and gamma is the natural linewidth of the atomic transition from $\left|e\right\rangle\rightarrow\left|g\right\rangle$. In the present experiment, the relatively large length of the cavity decreases the finesse, to the point where our single atom-cavity coupling $g_0 <$($\kappa,\gamma$). Nevertheless, when a large number of atoms are present in the cavity mode volume, as in the case when we make a MOT within the cavity, the collective coupling of $N$ atoms to the cavity mode results in the normal mode splitting of the atom-cavity resonance by
$2 g_{eff}=2 g_0 \sqrt{N}$~\cite{gsa1,carmichael2,mossberg1},which is detected in our experiment when $N > 1000$ atoms of $^{85}$Rb present in the mode volume. The strong coupling formalism is applied within the Tavis-Cummings model~\cite{Tavis1968,Tavis1969} of a two level atom interacting with the quantized electromagnetic field. Real atoms have many levels and therefore the transitions from excited electronic states are not necessarily to a single state.\\
\begin{figure}[t]
\centering
\includegraphics[width=15 cm]{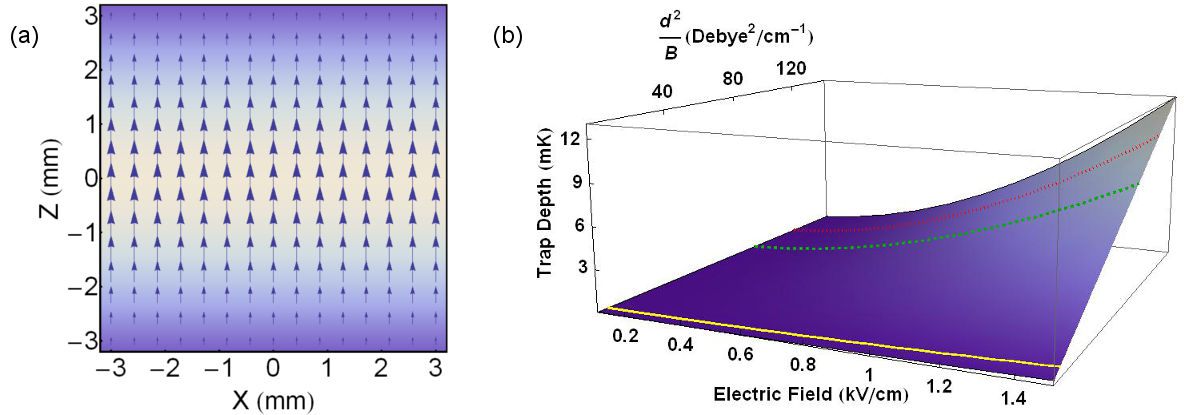}
\caption{(a) Shows the electric field in the XZ plane when the inner two wires are $+1$kV and $-1$kV respectively, which can be tuned for precision measurements in uniform electric fields. (b) shows the variation of the trap depth for polar molecules with the electric field and $\frac{d^{2}}{B}$. The yellow (solid), green (dashed)and red (dotted) lines represents the trap depths for RbK, RbCs and KCs molecules respectively, in their rovibrational and electronic ground state. In both figures, light to dark backgrounds indicates the  magnitudes of the quantities being plotted, from large to small.}
\label{fig:molecule}
\end{figure}
In Fig.~\ref{fig:AtomCavity}(c), we demonstrate the collective normal mode splitting on the transition from 5$S_{1/2}$, $F=3\leftrightarrow 5P_{3/2}$, $F'=2,3,4$ states of $^{85}$Rb. For the experiment the MOT atoms are prepared in the 5$S_{1/2}$, $F=3$ ground state and then interrogated with a very weak probe light through the cavity mode, scanning across the atomic resonances while the cavity is tuned to the corresponding atomic transition. A subclass of dipole allowed transitions are nearly closed in a $^{85}$Rb atom, corresponding to the J-C model. In Ray et. al.~\cite{Ray2013b}, we demonstrate the normal mode splitting for both closed and open transitions. In the case of open transitions we find that the atoms need to be repumped from the competing ground state while being probed, failing which the split does not manifest. Interestingly there is little polarization dependence of the energy split, when the atom-cavity system is probed in presence of the gradient magnetic field. The extent of the normal mode split proportional to the dipole matrix element between the states being accessed. Given that the presence of MOT gradient magnetic fields and the electric fields of the ion trap has no measureable effect on the normal mode split, makes this a potentially attractive system for probing complex mixtures interacting with cold atoms in the cavity mode. 

\section{Other features and prospects with the hybrid trap}
In principle all the experiments described so far would also be possible with a segmented linear Paul trap with an axial cavity. Where the thin-wire structure comes into its own is in its ability to allow strong constant fields about the trap volume. Fig~\ref{fig:molecule}(a) demonstrates that a relatively large electric field can be applied to trapped ensembles of atoms and molecules, which could then be probed by the cavity to make precision measurements on cold gases with tunable electric fields~\cite{Agarwal2004}. A further prospect with this electrode structure is the trapping of polar molecules, which can be done by a straight forward implementation of a gradient field trap, as demonstrated by Kleinert et. al.~\cite{Kleinert2007}. Fig.~\ref{fig:molecule}(b) shows the trap depth for heteronuclear diatomic molecules as a function of electric field and $d^{2}/B$, where $d$ is the dipole moment and $B$ is the rotational constant. Trap depths of RbK, RbCs and KCs molecules in their rovibrational and electronic ground states are marked on this figure, where the molecular constants for the calculations are taken from~\cite{Ni2008, Aymar2005, Allouche2000,Ferber2008}. In addition there are two overlapping ion traps in this configuration (linear and modified spherical) which can be used flexibly to many purposes. The experiment supports many atomic species and allows the production, manipulation and detection of molecular ions. The cavity can be utilized to define standing wave dipole traps for the neutral cold atoms and molecules, which can be overlapped with the trapped ions. Laser coolable ions of Ca$^+$ can be used so that the optical detection of ions can be used for specific experiments.  

\section{Conclusions} We have presented a novel design of ion trap combined with overlapped cold atom traps, and the mixture is contained in a cavity, which is a trap for resonant light. To our knowledge, this is the first realization of all these simultaneous and interacting elements of atomic physics. Molecules are also trappable in this apparatus, making it a very versatile experimental apparatus. The experiments discussed, which demonstrate the power and the utility of the apparatus are merely the tip of the iceberg, with respect to the potential of the system for doing mixed species experiment.  

\section{Acknowledgement} The experiments discussed above have evolved directly from the work of K. Ravi, Seunghyun Lee, Arijit Sharma and Guenter Werth, whose contributions are gratefully acknowledged by the Authors. Technical support by S. Sujatha and N. Narayanaswami and the RRI workshops is greatfully acknowledged.


\end{document}